# An *in-situ* X-ray and neutron diffraction investigation of Bi-2212 in multifilamentary wires during thermal treatment

Alberto Martinelli,[1] Emilio Bellingeri,[1] Alessandro Leveratto,[1] Luca Leoncino,[1] Clemens Ritter,[2] and Andrea Malagoli[1, *]

[1]*CNR-SPIN, Corso Perrone 24, 16152, Genova - Italy* [2]*Institut Laue-Langevin, 71 Avenue des Martyrs, 38042 Grenoble, Cedex 9 - France* (Dated: July 30, 2018)

* andrea.malagoli@spin.cnr.it

Significant insights for critical current density ($J_c$) improvement in Bi-2212 superconductor wires can be obtained by an accurate analysis of the structural and microstructural properties evolving during the so-called partial-melt process, a heat treatment needed to improve grain connectivity and therefore gain high $J_c$. Here, we report an *in-situ* analysis by means of synchrotron X-ray and neutron diffraction performed, for the first time, during the heat treatment carried out with the very same temperature profile and reacting oxygen atmosphere in which the Bi-2212 wires are usually treated for practical applications. The obtained results show the thermal evolution of the Bi-2212 structure, focusing in particular on texturing and secondary phases formation. The role of the oxygen is discussed as well. Hence, the present investigation marks a significant advance for the comprehension of the phenomena involved in the wire fabrication process and provides useful insights for the process optimization as well.

## I. Introduction

Scientific applications involving magnet construction for the energy upgrade of Large Hadron Collider (LHC), for fusion (ITER), or for Nuclear Magnetic Resonance (NMR) magnets require superconducting materials capable of generating magnetic fields above 20 T, once wound as magnets [1] [2]. High temperature superconductors such as $YBa_2Cu_3O_{7-x}$ (YBCO), $(BiPb)_2Sr_2Ca_2Cu_3O_{10-x}$ (Bi-2223) and $Bi_2Sr_2CaCu_2O_{8+x}$ (Bi-2212) show flat $J_c$ behaviour up to fields above 30 T due to $H_{c2}$ exceeding 100 T [3]. Among these, Bi-2212 wires exploit a very flexible technology: they can be processed as round conductor with fine and twisted filaments arranged in multiple architectures. The metallic sheath has a high thermal conductivity and there is no need for any diffusion barrier. However, the significantly low critical current density ($J_c$) of Bi-2212 wires has been the bottleneck for long time to their application for magnet construction. For several years, research works aimed to improve such transport properties were focused on the analysis of the Bi-2212 phase formation trying to reveal the relationship between processing, microstructure and superconducting properties. Despite of that, recent reports [4] [5] [6] [7] revealed that the internal gas pressure generated during the heat treatment produces the formation of bubbles inside the wire and the density of the superconducting filaments is reduced. Such a phenomenon has been recognized as the main hindrance for $J_c$ enhancement in Bi-2212 wires. The application of a gaseous over pressure (50–100 bar; OP) during the heat treatment has been proposed to increase the density of the Bi-2212 filaments and thus $J_c$ [8]. The samples obtained by the OP process show engineering critical current density ($J_e$) values of 800–900 A/mm$^2$ at high fields, well above the application require- ments ($J_e$=400–600 A/mm$^2$). As an alternative route to the OP process, an innovative mechanical deformation has been proposed by our group [9] [10] [11] with the aim to realize dense wires with superconducting properties that could satisfy the application requirements through a more attractive process for an industrial scaling-up. In such a context, significant insights for $J_c$ improvement in this material can still be obtained by an accurate analysis of the evolution of the structural and microstructural properties during the so called partial-melt process (PMP), a heat treatment needed to improve grain connectivity and therefore gain high $J_c$. Such a process involves the heating of Bi-2212 powders contained into the Ag sheath above the peritectic temperature (∼ 882°C); as a result, Bi-2212 decomposes into a melt plus solid $(SrCa)_{14}Cu_{24}O_{4x}$ (14:24 AEC) and $Bi_2(SrCa)_4O_x$ (2:4 CF) phases. During subsequent slow cooling this mixture reacts, forming well connected Bi-2212 phase. The final annealing and subsequent cooling at room temperature lead to an oxygen over-doping, enhancing the transport properties at high fields [12]. The investigation of the phase assemblage in Ag-sheathed Bi-2212 wires and its evolution during thermal treatment has been the subject of several studies. Zhang and Hellstrom [13] found that the phase assemblage forming after the incongruent melting of Bi-2212 is strongly dependent on the $O_2$ partial pressure in the reacting atmosphere. Recent works showed an extensive formation of Bi-2201 ($Bi_2Sr_2CuO_6$) intergrowths within Bi-2212 grains [14]. The presence of Bi-2201 and secondary phases in fully reacted wires can represent a limit in reaching very high $J_c$ even in very dense filaments. All of these studies were performed using SEM-EDX analysis on fully reacted wires and/or samples quenched from high-temperature to room temperature. A first attempt to an *in-situ* analysis was performed by C. Scheuerlein et al [15] but here, for the first time, the phases and structure evolution have been analysed during the complete heat treatment, where each step has been accurately monitored in terms of heating/cooling rate, temperature and duration. Actually, neither the processes occurring during each step of the heat treatment nor the role of the oxygen pressure are yet completely understood. We recently applied, for the first time to the Bi-2212 superconductor, a model based on the proximity effect to describe the nature of the grain boundaries through the behavior of $J_c$ as a function of the temperature in samples with different oxygen content in the overdoping regime [16]. In the present work, we report the results of an *in-situ* investigation of Bi-2212 wires, carried out throughout the thermal treatment performed in the very same conditions usually applied in the production process. In particular, the reported results illustrate how the Bi-2212 and the secondary intergrowth phase Bi-2201 evolve during the thermal treatment as well as the major role played by the oxygen activity in the reacting atmo- sphere for crystallization and texturing. The reported results are hence important not only for the comprehension of the phenomena involved during the wire fabrication pro- cess, but also because they can provide useful insights for the optimization of the process itself.

## II. Experimental

The investigated samples were prepared using the powder-in-tube method (PIT), start- ing by filling an Ag tube with outer (OD) and inner (ID) diameters of 15 and 11 mm, respectively, with Nexans granulate powder of composition $Bi_{2.16}Sr_{1.93}Ca_{0.89}Cu_{2.02}O_x$. Af- ter drawing, the obtained monofilamentary wire is hexagonally-shaped, cut in 331 pieces and restacked in a 18/15 mm (OD/ID) Ag/Mg alloy tube. It is then cold worked with a proper alternation of drawing and groove-rolling steps. Finally, a square wire is obtained, having a $0.7 \times 0.7$ mm$^2$ cross-section, an average size of the single filament of about 20–25 μm and a superconducting fill factor of about 22%. In the above described process, the new concept for the densification of powders before the partial melt process is used [9], namely a proper alternation of drawing and groove-rolling steps that drastically reduces the porosity and leads to a powder density inside the filaments much higher than that obtained by the standard process with just a drawing deformation. Figure 1 shows the cross-section of a multifilamenatry wire.

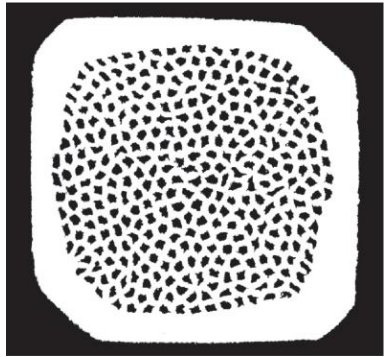

FIG. 1. Cross-section of the multifilamentary wire: $0.7 \times 0.7$ mm$^2$ in size and 331 filaments.

The heat treatment optimized by Oxford Superconducting Technology (OST) for their commercial Bi-2212 wires, commonly used for conductor and magnet development, has been selected as a reference during our experiments (Figure 2; TT1). The sample wire, after being processed with such thermal treatment, showed a critical current density $J_c = 2300$ A/mm$^2$ at a temperature of 4.2 K and an applied magnetic field of 5 T. In the same condition of temperature and magnetic field, the wire processed in air with the same thermal treatment TT1 showed a $J_c = 300$ A/mm$^2$. Such values are in good agreement with the values reported in literature [16].

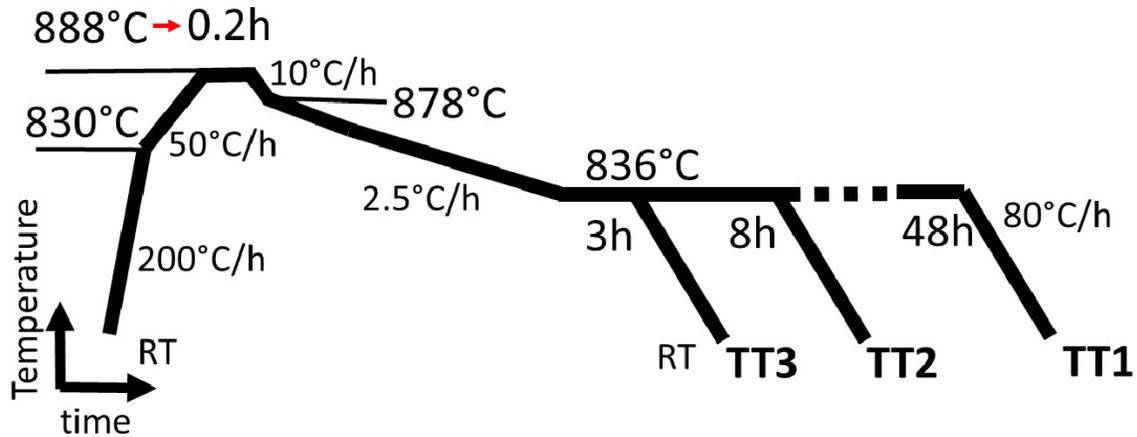

FIG. 2. Scheme (not in scale) of the thermal treatment profiles (temperature versus time) used during experiments. The three treatments have the same temperature profile from the beginning until the plateau at 836 °C, whereas differ in the duration of the plateau itself.

Due to the limitations dictated by the available beam time at the large scale facili- ties where experiments were carried out, some changes were necessary, as hereafter de- scribed. The *in-situ* powder diffraction analyses throughout thermal treatements were carried out using X-ray radiation at the ID11 beamline of the European Synchrotron Radiation Facil- ity (ESRF; Grenoble, France; wavelength λ= 0.14082 Å) and neutron radiation at the high- intensity D20 diffractometer (λ = 2.4162 Å) of the Institute Laue Langevin (ILL; Grenoble, France). In both experiments, the samples were placed in a furnace mounted directly on the neutron or X-ray beamline. Figures 3 and 4 illustrates the experimental set-up assembled at ESRF and ILL, respectively.

The XRD analysis at ID11 was carried out using a tubular furnace in an experimental set- up similar to the one used by Flukiger et al [17]. The quartz tube was closed by two kapton plugs and suitably shaped radiation screens were realized; in this way, the in- cident and diffracted radiation can be transmitted without interference across a takeout angle of about 3 degrees. Single multifilamentary wires were then reacted under both an O$_2$ flux and an air flux using the thermal treatments TT2 and TT3, respectively (Figure 2). XRD patterns were collected in transmission geometry; a two dimensional detector was used and images were integrated along the azimuthal direction in order to obtain q-scan type measurements for the study of the phase evolution.

In order to optimize the available beam-time, the thermal treatments were arranged as follows: the plateau step has been reduced at 3 h and 8 h instead of 48 h for the treatments.

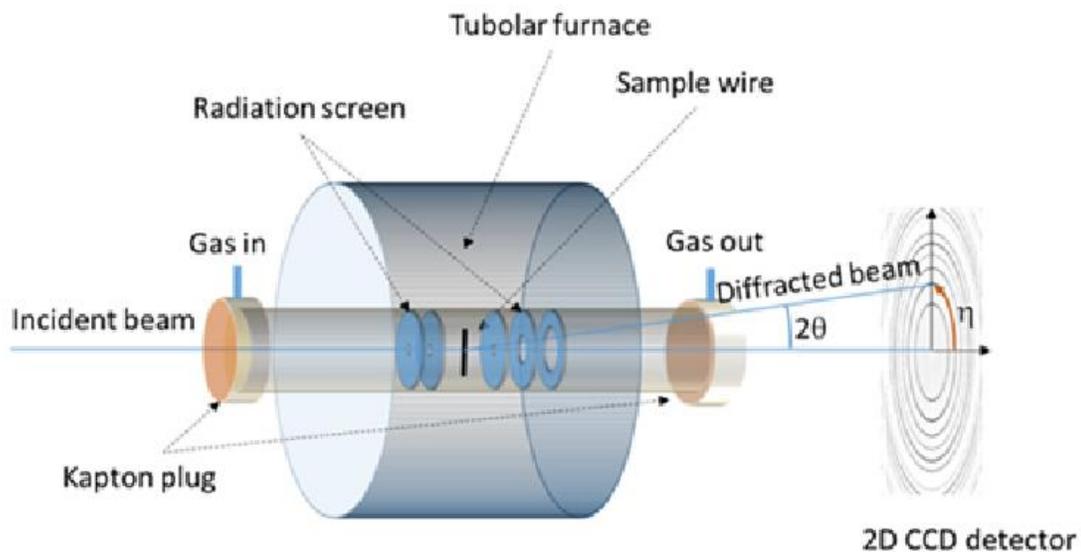
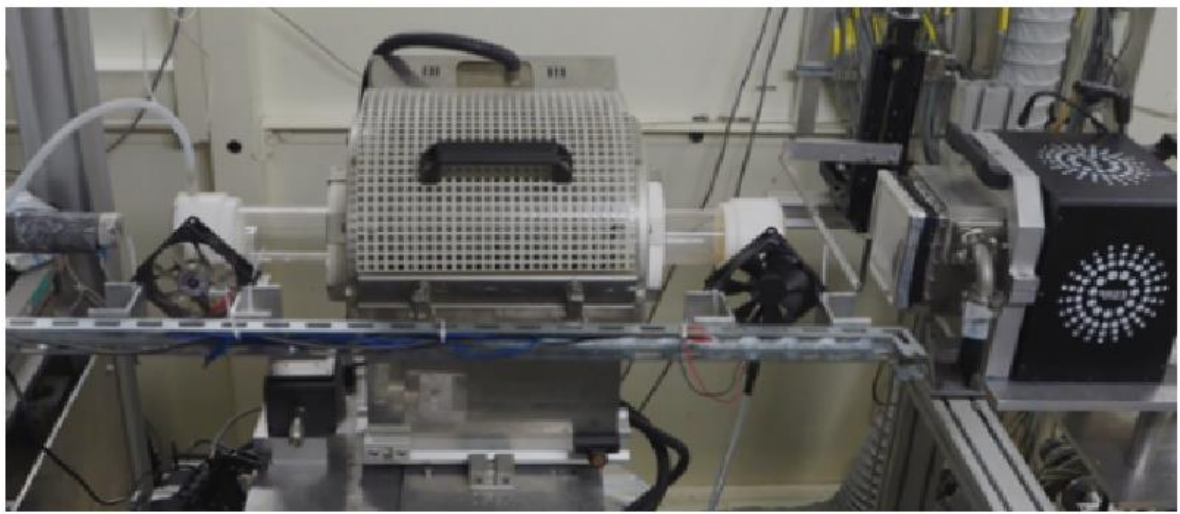

FIG.3. Experimental set-up employed during XRD data collection at the ID11 beamline of ESRF. The 2θ Bragg angle and η azimuthal angle are indicate.

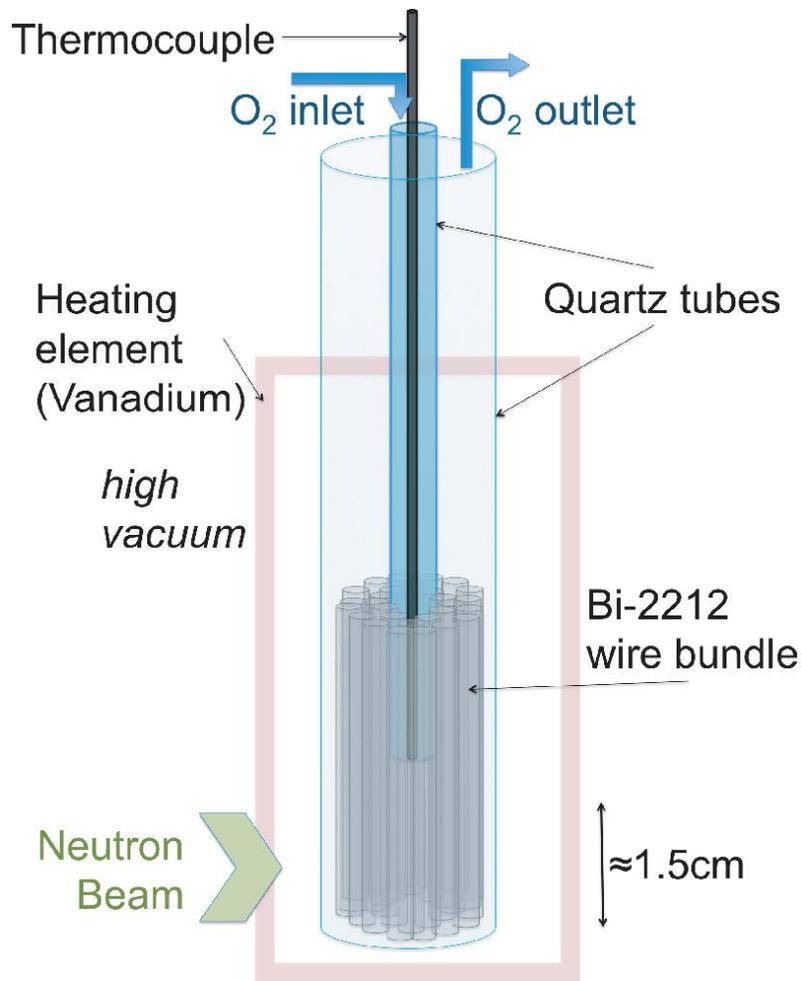

FIG.4. Experimental set-up of the samples used in the neutron diffraction experiment at ILL.

in air (TT3) and $O_2$ (TT2), respectively. In order to obtain a more complete analysis, data collection at room temperature has been performed on samples that previously under- went the whole heat treatment TT1 in our laboratories (fully reacted samples) in both air and $O_2$ atmosphere. At this scope, data collection was also carried out at room tempera- ture without the heating furnace, thus providing diffraction patterns with wider angular ranges. In order to estimate the weight percentage of the coexisting Bi-2212 and Bi-2201 phases, these data were fitted using the Rietveld method.

The neutron powder diffraction (NPD) patterns were collected on a bundle of multifil- amentary wires under an $O_2$ flux and using the thermal treatment TT2. In the neutron experiment an ILL vanadium furnace was used; the sample was inserted inside a quartz tube and hold at the optimal position by a second smaller quartz tube having the addi- tional function of $O_2$ inlet. A Pt/PtRh thermocouple in contact with the wire bundle was used for the feedback control of the temperature [18].

The accuracies of both furnaces used during the XRD and NPD experiments are ±0.5 °C on the length of the samples themselves. Both NPD and XRD data were used to evaluate the phase evolution as well as the thermal dependence of microstructural properties of Bi-2212. Noteworthy, most of the radiation scattered by the internal filaments is absorbed by the external Ag sheath. For this reason, structural refinements cannot be carried out; nonetheless, Le Bail fittings (profile match- ing) can be reliably performed. At this scope fitting was carried out using the program FULLPROF [19] using a file describing the instrumental resolution function and refin- ing the scale factor; the zero point of detector; the background; the unit cell parameters; selected profile parameters; the preferred orientation parameters. The microstructural properties of the Ag sheath were inspected by means of the Williamson-Hall method [20]; in particular, in the Williamson-Hall plots a straight line passing through the origin and all the points of the different orders of the same reflection is obtained, whose slope provides a qualitative evaluation of the microstrain contribution. In addition, the texture was evaluated by analyzing the azimuthal distribution of the XRD peak intensity.

After the thermal treatment 1, Bi-2212 powders extracted from the wires were analysed by X-ray powder diffraction analysis (XRPD; PANalytical X'Pert PRO; CuK$_\alpha$). The crystal structures were refined according to the Rietveld method [21]; refinements were carried out using a file describing the instrumental resolution function. In the final cycle, the following parameters were refined: the scale factor; the zero point of detector; the background (parameters of the 5$^{th}$ order polynomial function); the unit cell parameters; the atomic site coordinates not constrained by symmetry; the atomic displacement parame- ters; the anisotropic strain parameters; the preferred orientation parameters.

### III. Results

**A. Structural properties of Bi-2212**

Despite its attractive physical and technological properties, a comprehensive inves- tigation of the structural properties at high temperature of the Bi-2212 phase has not yet carried out, to our knowledge. Structural studies performed at room temperature report both a tetragonal (space group *I4/mmm*) [22] as well as an orthorhombic (space group *Fmmm*) [23] symmetry. Rietveld refinement using X-ray powder diffraction data collected on the starting Bi-2212 powders suggests a tetragonal symmetry; in fact, no ev- ident peak splitting of the tetragonal 110 diffraction line into the orthorhombic 200 and 020 ones can be appreciated. Figure 5 shows a representation of the layered structures of both Bi$_2$Sr$_2$Ca$_1$Cu$_2$O$_{8+x}$ (Bi-2212) and the closely related Bi$_2$Sr$_2$CuO$_6$ (Bi-2201) compound. Under suitable conditions, this structural correlation favours the intergrowth of Bi-2201 layers in melt-processed Bi-2212.

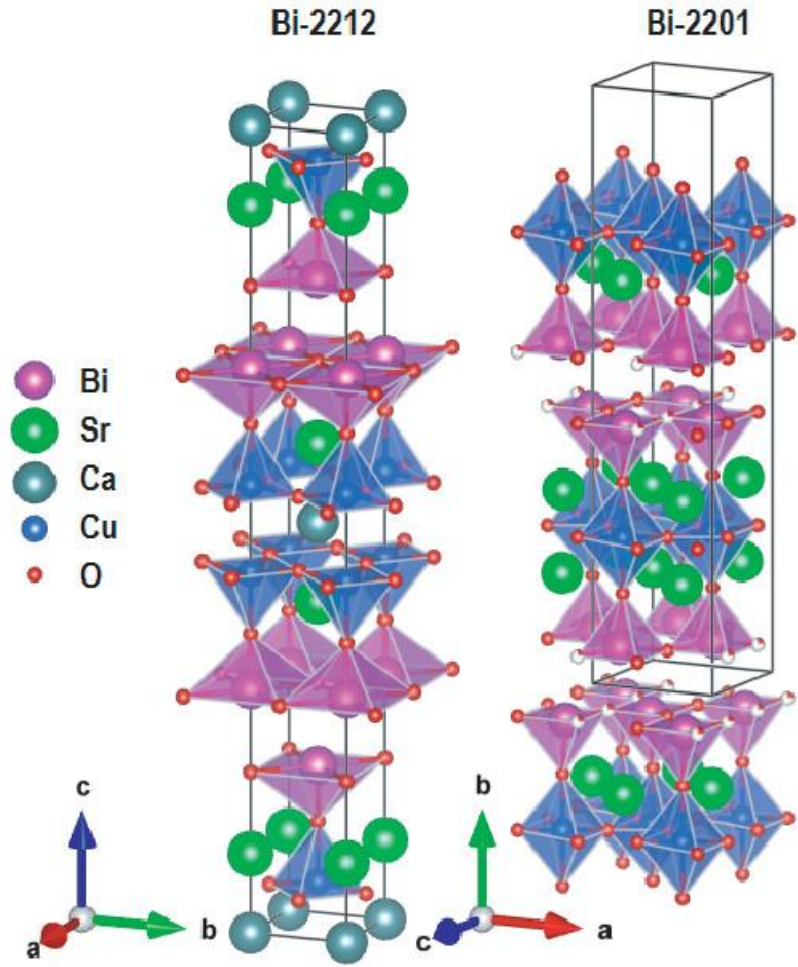

FIG.5. Representation of the crystal structures of tetragonal Bi-2212 and Bi-2201.

Figure 6 shows the comparison of the XRPD patterns for tetragonal Bi-2212 and Bi- 2201 in the Q-range mainly inspected in the present work. For the sake of clarity, the position of these Bragg peaks serve as a reference, but can be affected by not-negligible shifts, depending on both the experimental set-up (zero shift of the diffractometer), tem- perature (thermal expansion) and chemical composition (mainly O content).

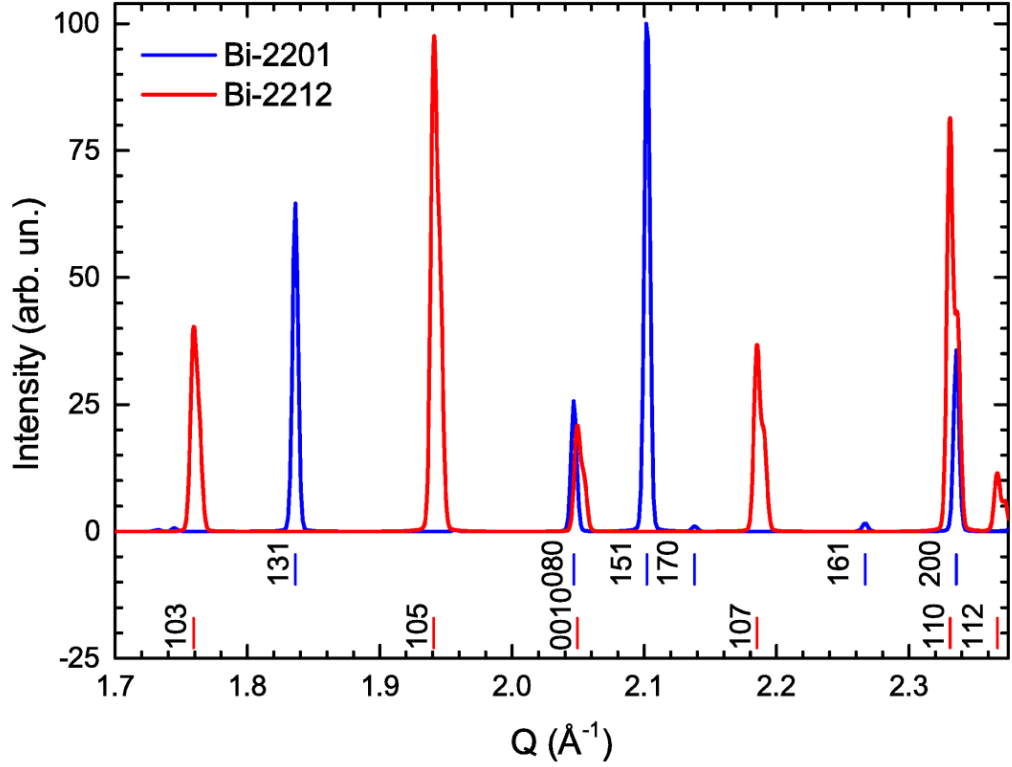

FIG. 6. Comparison of the calculated X-ray powder diffraction patterns for Bi-2212 and Bi-2201 (selected Q-range); vertical bars represent the position of the Bragg peaks (indexed).

### B. Structural refinement of Bi-2212 after thermal treatments

Table I lists the structural data of Bi-2212 extracted from the multifilamentary wires after the thermal treatments TT1 in air and $O_2$ (after Rietveld refinements using the labo- ratory XRPD data).

Rikel and Hellstrom [24] pointed out that the density of the Bi-2201 intragrowth layers (f) increases with the increase of the texturing degree; moreover, these authors reported a correlation between the intensity ratio of the 0 0 8 and 0 0 10 diffraction lines with the average <f(1-f)> quantity. By applying this correlation, the density of the Bi-2201 layers intragrown within the Bi-2212 phase can be approximately estimated, being ~6.5% at maximum in the $O_2$ treated sample and negligible in the air treated sample. From these data, it can be concluded that the degree of texturing in our samples is significantly increased in the $O_2$ sample, a result that is confirmed by the *in-situ* analysis data reported in III D.

|  | | air | $O_2$ |
|---|---|---|---|
| a (Å) | | 3.8202(1) | 3.8242(1) |
| c (Å) | | 30.8076(1) | 30.7685(1) |

| atom | Wyckoff site | x | y | z | x | y | z |
|---|---|---|---|---|---|---|---|
| Bi | 4e | 0 | 0 | 0.3019(1) | 0 | 0 | 0.3012(1) |
| Sr | 4e | 0 | 0 | 0.1094(1) | 0 | 0 | 0.1109(1) |
| Ca | 2a | 0 | 0 | 0 | 0 | 0 | 0 |
| Cu | 4e | 0 | 0 | 0.4485(1) | 0 | 0 | 0.4509(1) |
| O(1) | 8g | 0 | 1/2 | 0.0693(1) | 0 | 1/2 | 0.0713(1) |
| O(2) | 4e | 0 | 0 | 0.2215(1) | 0 | 0 | 0.2304(1) |
| O(3) | 4e | 0 | 0 | 0.3738(1) | 0 | 0 | 0.3762(1) |
| $R_{Bragg}$(%) | | | | 16.8 | | | 14.9 |

TABLE I. Structural parameters of Bi-2212 thermal treatments TT1; space group *I4/mmm*).

## C. Thermal evolution of the phase assemblage in the filaments

The thermal treatments at which the multifilamentary wires are submitted can be roughly divided into 5 main steps: 1) rapid heating up to 888°C, above the melting point of the Bi-2212 phase; 2) melting of the Bi-2212 phase; 3) texturing of the Bi-2212 phase on slow cooling; 4) plateau step at 836 °C; 5) cooling down to room temperature. Figure 7 shows the XRD data from the multifilamentary wires collected during over the samples (laboratory XRPD data collected at 300 K after melting step. In both samples the Bi-2212 phase decomposes into the Bi-2201 phase (peaks at ~1.80, ~2.07 and ~2.32 Å$^{-1}$) at ~610 °C (~2.83 h). The observed intensities of the Bi-2201 Bragg peaks indicate that texturing is negligible at this stage, conversely to what observed during the texturing step (*vide infra*). As the temperature further in- creases, the Bi-2201 phase decomposes forming a novel Bi-2212 phase approximately at ~850 °C (~4 h). Then, the Bi-2212 phase melts at the peritectic temperature (~882°C).

Upon melting, during the slow cooling step, some diffraction peaks can be detected in both air and $O_2$ treated filaments, but the correct identification of the corresponding phases is extremely challenging. Nonetheless, some clues can be gained: the occurrence of the $Bi_2(SrCa)_4O_x$ (2:4 CF) phase can be mainly deduced from the presence of a main peak centred at ~2.11 Å$^{-1}$ (Figure 7) and fainter ones at ~1.21 Å$^{-1}$ and ~3.65 Å$^{-1}$. Figure 8 shows the phase evolution as recorded by NPD in about the same Q-range as the X-ray data shown in Figure 7. It should be noted here that due to the different scattering lengths of the elements in X-ray and neutron diffraction the intensities of the Bragg peaks can be very different. The peak at ~2.11 Å$^{-1}$ evidences the presence of the 2:4 CF phase again; unfortunately, the crystal structure of this phase at high temperature has not yet been resolved and hence it is challenging to discriminate its diffraction lines in a NPD pattern. Nonetheless, on account of their similar behaviour, also the peaks at ~2.20 Å$^{-1}$, ~2.27 Å$^{-1}$ and ~2.34 Å$^{-1}$ can be presumptively ascribed to the 2:4 CF phase. In this con- text, it is worth to note that the $Bi_6(Sr_{8-x}Ca_{3+x})O_{22}$ (-0.5 ≤ x ≤ 1.7) compound reported by Luhrs et al. [25] likely corresponds to 2:4 CF phase, after comparison of the respective compositions and cell parameters.

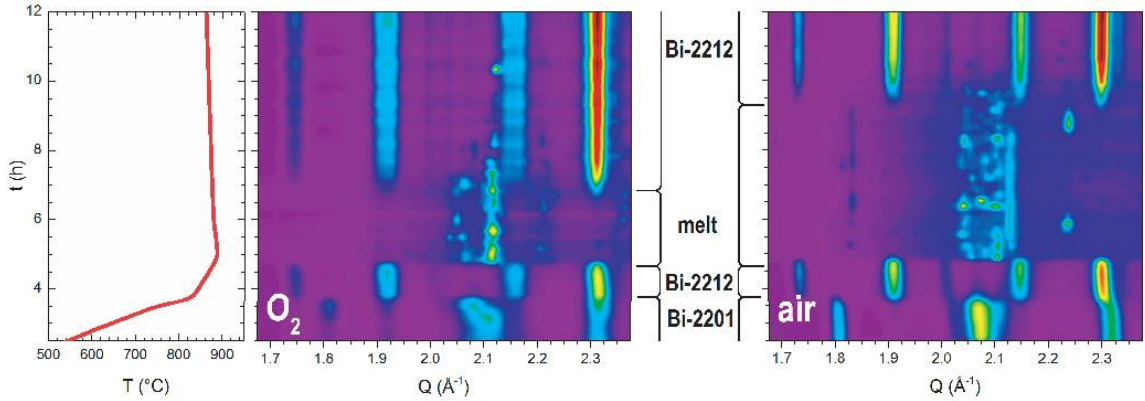

FIG. 7. Phase evolution in the multifilamentary wires during the first 12 hours of the thermal treatment in $O_2$ (on the left) and air (on the right); selected Q-range of XRD data.

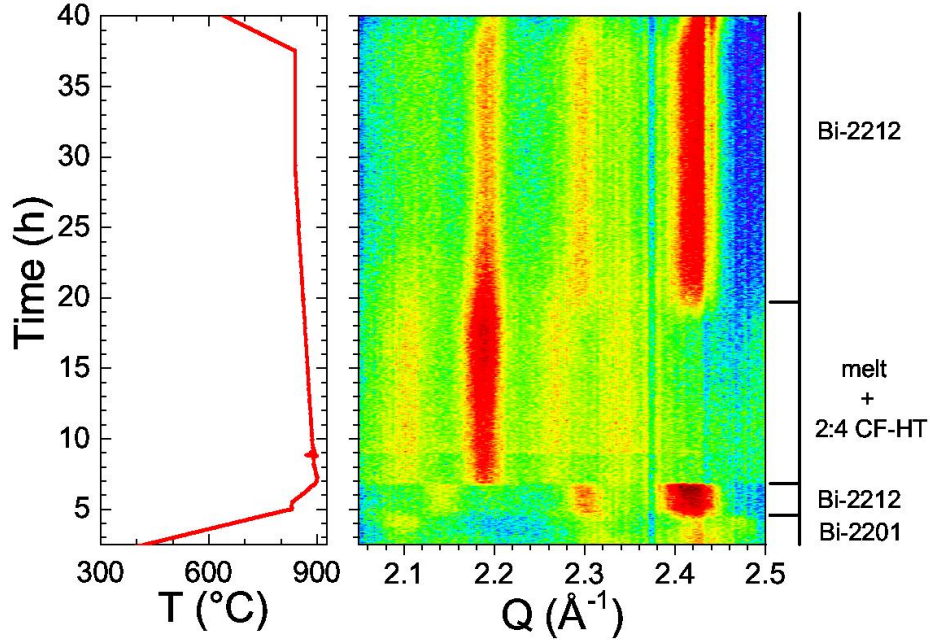

FIG. 8. Phase evolution in the multifilamentary wires (thermal treatment TT2 in $O_2$); selected Q-range of NPD data.

Concerning the air treated sample, relatively strong diffraction lines are observed in the XRD patterns at 2.04, 2.07 and 2.10 Å$^{-1}$ during the melting step, consistent with pre- vious investigations [15] [26]. In these works these peaks were ascribed to the presence of the high temperature polymorph of the $Bi_2(Sr_{4-x}Ca_x)O_7$ phase. Actually, a closer inspec- tion reveals their different trends over time (Figure 9), suggesting that they are produced by distinct phases. The peak observed at 2.20 Å$^{-1}$ in the XRD patterns could be related to the $(SrCa)_{14}Cu_{24}O_{41}$ phase (14:24 AEC); its intensity decreases with time, whereas during the slow cooling faint peaks grow at ~0.95, 2.31 and 2.36 Å$^{-1}$; likely due to the formation of $(SrCa)_2CuO_3$ (2:1 AEC). This behaviour is consistent with previous analyses carried out under an $O_2$ partial pressure ranging between 0.40 and 1.00 atm [13], were it was found that Bi-2212 melts incongruently forming the 2:4 CF and the 14:24 AEC phases, this last being replaced later by the 2:1 AEC phase.

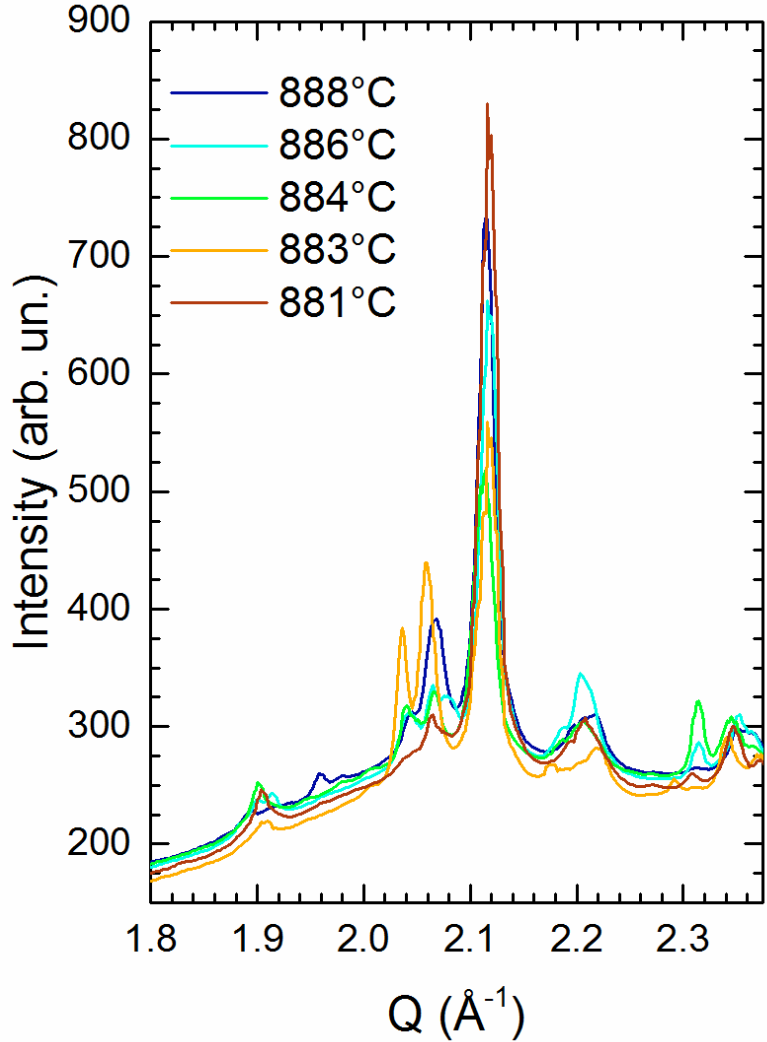

FIG. 9. Evolution of the phase relationship with time within the melted filaments ($O_2$ treated sample); XRPD diffraction patterns are collected every 10 min.

### D. Growth and texturing of the Bi-2212 phase from the melt

On further cooling the Bi-2212 phase starts to grow, representing the main phase in both samples. The crystallization of the Bi-2212 phase from the melt is remarkably ruled by the reacting atmosphere. In the $O_2$-treated sample the Bi-2212 phase starts to crystal- lize after ~6.58 h (876 °C), exhibiting an evident preferred orientation along [110] (Figure 10). This result is consistent with previous studies carried out on round wires [27] and melt-cast processed polycrystalline tubes of Bi-2212 [28], where a preferred orientation along [100] or [010] was detected by assuming an orthorhombic structural model (in the orthorhombic structural model the a and b axes are rotated by 45° and these lattice parameters are a factor of √2 longer compared with the tetragonal *a* and *b* axes). Conversely, it starts to grow only after more than 9 hours (869 °C) in the air treated sample; also in this sample the preferred orientation along [110] is evident, but in this case the remaining diffraction peaks exhibit noticeable intensities, thus indicating a lower degree of textur- ing.

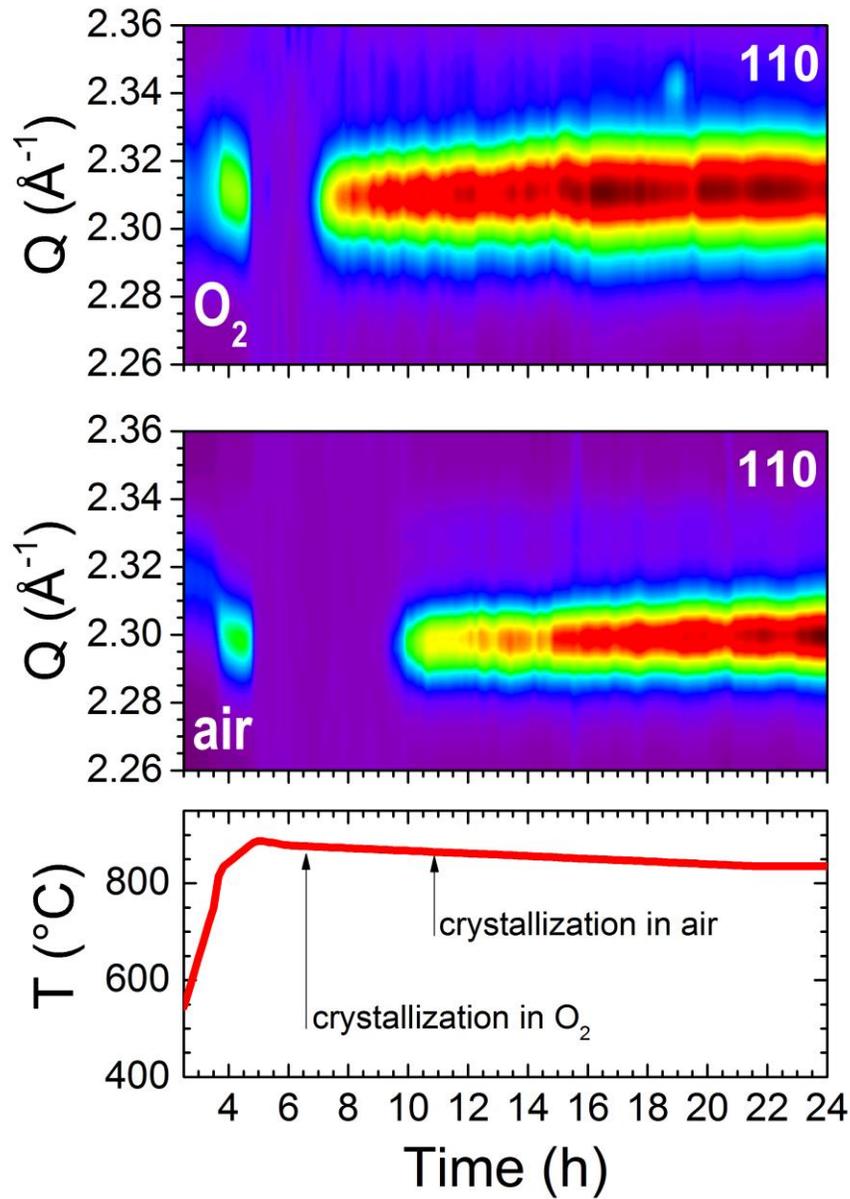

FIG. 10. Time dependence of the 110 diffraction line of the Bi-2212 phase during the thermal treatments in $O_2$ and air (XRPD data).

Figure 11 evidences the different time evolution of the integrated intensity for the tetragonal 110 peak of the Bi-2212 phase after its growth from the melt as obtained from the Le Bail fit for both $O_2$- and air-treated samples. For the $O_2$-treated sample, the intensity rapidly increases and then almost saturates, indicating that the Bi-2212 phase undergoes at first a fast growth. Later the intensity of the 110 peak remains approximately con-stant with time, suggesting that the amount of Bi-2212 does not change significantly. In the air-treated sample, where the growth of the Bi-2212 phase is substantially delayed, the formation of the Bi-2212 phase grows as well rapidly before entering a different regime where the reaction leading to the formation of the Bi-2212 is still taking place, although slowed down. On the basis of our data, it is not possible to ascertain if the reduced oxygen activity in the air-treated sample directly slows down the kinetic of formation of Bi-2212 or else favours the formation of secondary phases, whose growth competes with that of Bi-2212.

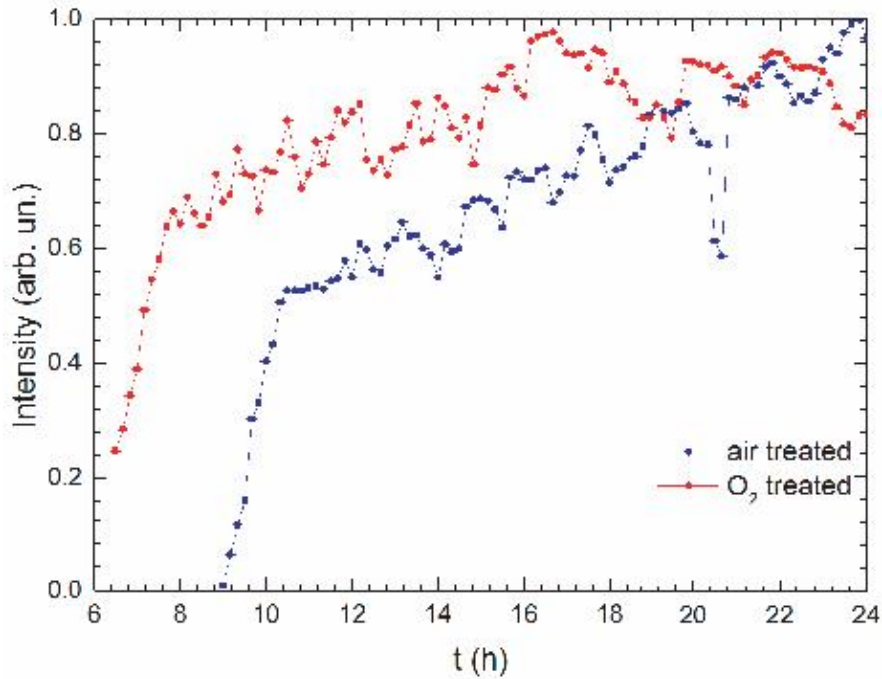

FIG.11. Time evolution of the integrated intensity for the tetragonal 110 diffraction line after the crystallization of the Bi-2212 phase from the melt (normalized data).

Figure 12 shows the texturing evolution of the Bi-2212 phase after crystallization from the melt during the thermal treatments in $O_2$ and air obtained from the refined texture parameter. Remarkably, more than 95% of the sample treated under an $O_2$ atmosphere is textured along [110] as the phase appears early after the melting; the sample treated in air displays as well a notable texturing along the same direction, but the amount of oriented phase is significantly reduced. Noteworthy, the refinement of the preferred ori- entation parameters indicates that in all samples the Bi-2212 crystals are characterized by a prevailing platy-like habit.

In order to deeply investigate the texture evolution, we directly studied the spatial distribution of the diffracted peak in the 2D acquired images (see Figure 13). Indeed, due to the strong texturing the angular distribution of the Bragg peaks is uneven and not equally distributed along Debye Scherrer rings (at constant Q) as expected for randomly oriented powders. Analyzing several Bragg peaks belonging to different {$hkl$} families and assuming a cylindrical symmetry along the axis of the wire, it is possible to extract from the images the orientation distribution of the superconducting grains inside the filaments. The position of the peaks gives a biaxial orientation in agreement with the results of Kametani et al. [27] with the *c*-axis laying in a plane normal to the RD (wire direction) and a less pronounced preferential orientation of the basal axes in the wire direction; analysing the azimuthal angular dispersion of Friedel pairs a FWHM of about 15° and 30° can be estimated for the two direction, respectively.

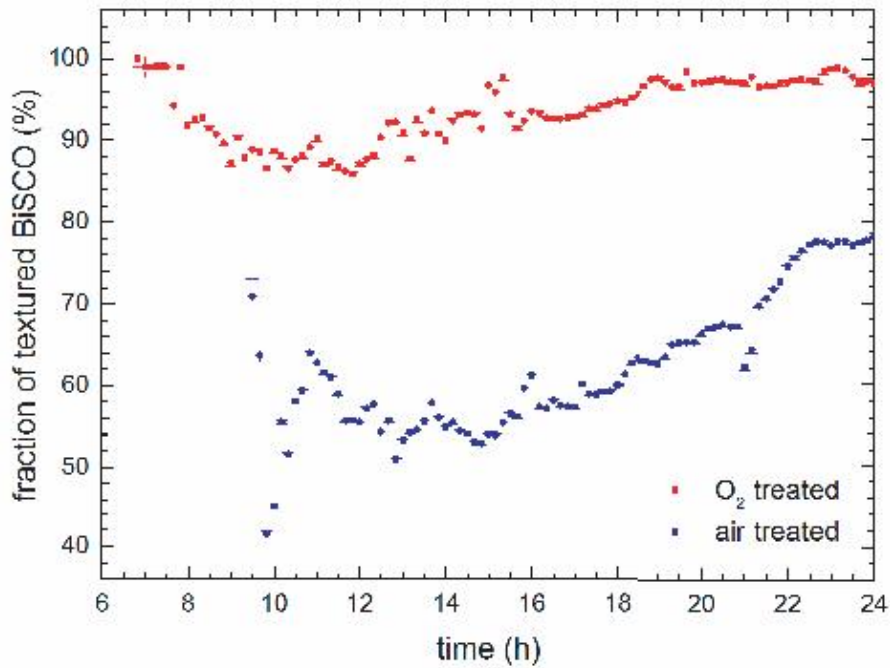

FIG.12. Evolution of the fraction of texture Bi-2212 phase as function of time during the thermal treatments in $O_2$ and air.

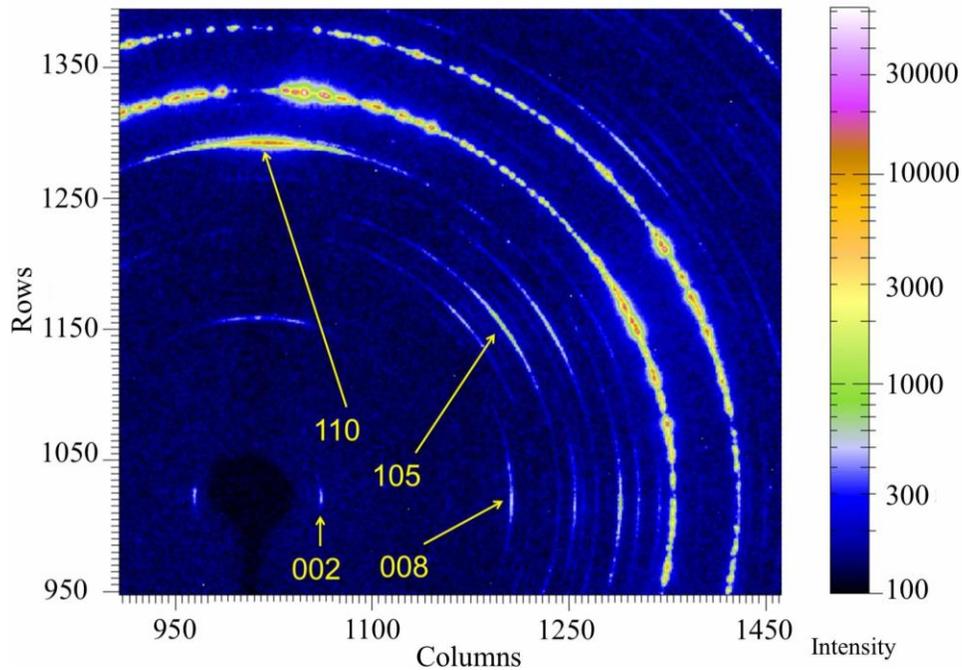

FIG.13. Acquired diffraction image showing the distribution of the spot.

We also analysed the origin and the evolution of this texture during the thermal treat- ment of the wire. In figure 14 the intensity of the 008 (b) and 105 (c) peaks for the $O_2$ treated sample as a function of the azimuth angle (horizontal axis) and of the time (verti- cal axis) are reported. Following the evolution of these two peaks it is possible to evaluate the progression of the *c*-axis texture and of the basal plane respectively. A small degree of texturing, with the platy-like grains oriented along the rolling direction (Figure 14b) (*c*-axis in the normal directions) but no in plane texture is present (Figure 14c) at the beginning of the thermal treatment. This effect is due to the cold working process that forces the grain to align with their largest dimension in the deformation directions, in a similar way on what happens in the Bi-2223 tapes fabrication where this effect is then amplified by the flat rolling. The strong texture appears instead immediately when the Bi-2212 phase starts to crystallize after the "melt" and no further significant effects are observed during the rest of the thermal treatment.

A similar behaviour is observed in samples prepared in air but some differences are present in the diffraction pattern, indicating a relatively more disordered structure with a more discrete distribution of the grain orientation (see Figure 14 and 15). The orientation is here again determined at the nucleation and crystallization point. Differently to the sample treated in $O_2$, it is, however, possible to observe some grains that during the first part of the cooling process rotate slightly (marked by arrows on the Fig 15c), but the overall original orientation acquired at the nucleation is kept during the process.

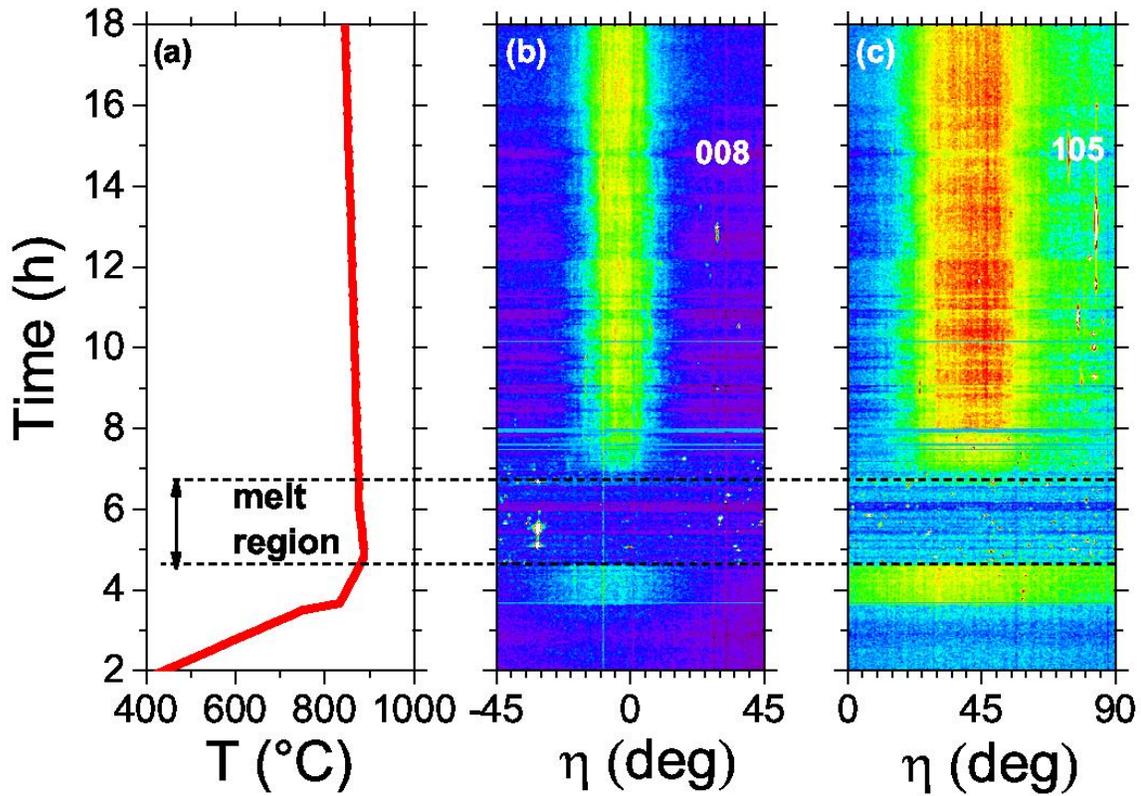

FIG. 14. Intensity of the 008 (b) and 105 (c) peaks as a function of the azimuth angle (horizontal axis) and of the time (vertical axis) during the thermal treatments in $O_2$ (a).

**E. Growth of the secondary Bi-2201 phase**

In parallel with the crystallization and texturing of the Bi-2212 phase also the growth of the Bi-2201 phase is observed following the melting step, as evidenced by the growth and evolution of the 151 peak with time (Figure 16). Using the Rietveld method, the weight percentage of the Bi-2201 phase can be determined to ~9.0(2.0)% in the air treated sample at the end of the thermal treatment, whereas in the $O_2$ treated sample this percentage is notably suppressed, being only 6.4(1.0)%. It is important to notice that in this case the considered Bi-2201 phase percentage corresponds to the one coexisting as a secondary phase with the main Bi-2212, not to the percentage which is intragrown within it (see III B). Remarkably, the amount of both Bi-2212 and Bi-2201 phases increases with time during the plateau step in the air-treated sample, suggesting that the formation processes of these phases compete for the same reactants.

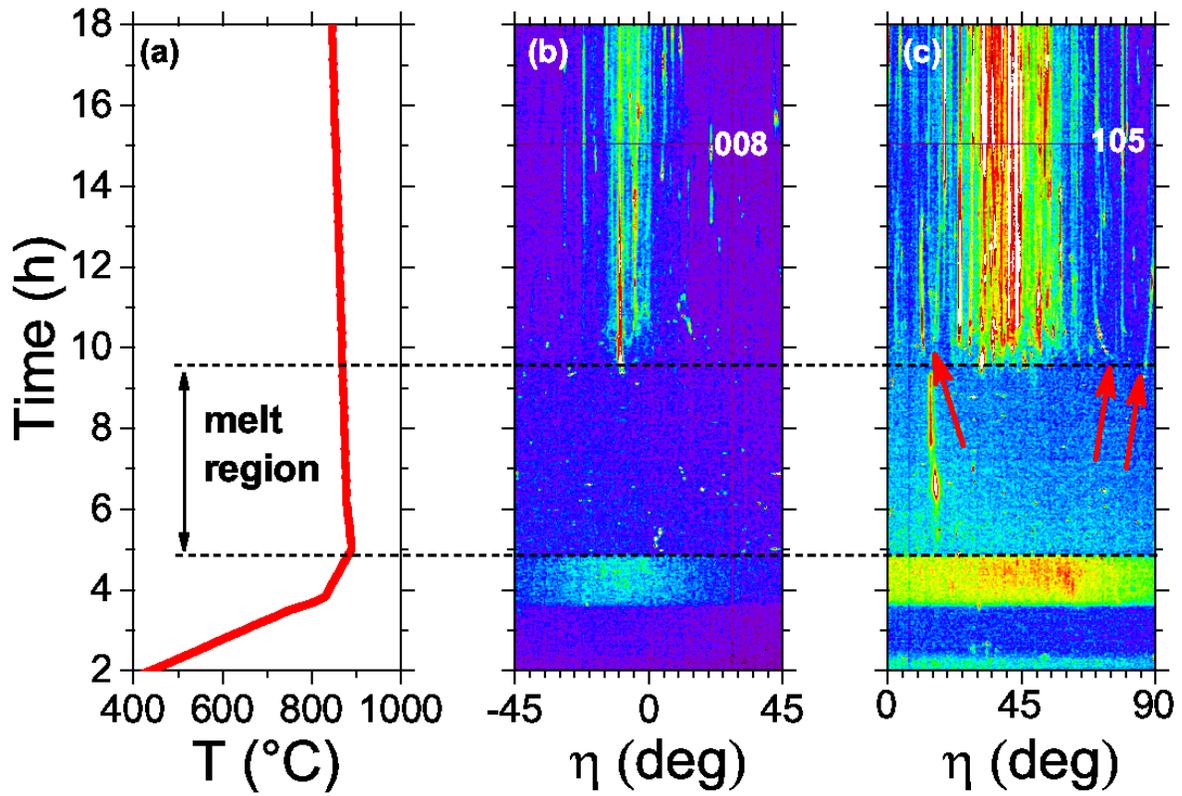

FIG.15. Intensities of the 008 (b) and 105 (c) peaks as a function of the azimuth angle (horizontal axis) and of the time (vertical axis) during the thermal treatments in Air (a).

### F. Analysis of the fully reacted samples

The synchrotron XRD data collected on the fully reacted samples has been fitted using the Rietveld method for estimating the weight percentage of the phases. The obtained re- sults are in good agreement with the afore-reported data, indicating a degree of texturing for Bi-2212 of 99% and 80% after the thermal treatments in $O_2$ and air, respectively (Fig- ure 17).

Several medium-to high intensity peaks pertaining to the Bi-2212 phase can be recognized in the air treated sample, although the 110 diffraction line (arrowed) exhibits the highest relative intensity, thus indicating a marked texturing. Differently, in the $O_2$ treated sample only the 110 diffraction line is characterized by a high relative intensity, all the other peaks displaying weak-to-undetectable intensities, thus confirming the higher degree of texturing in this sample.

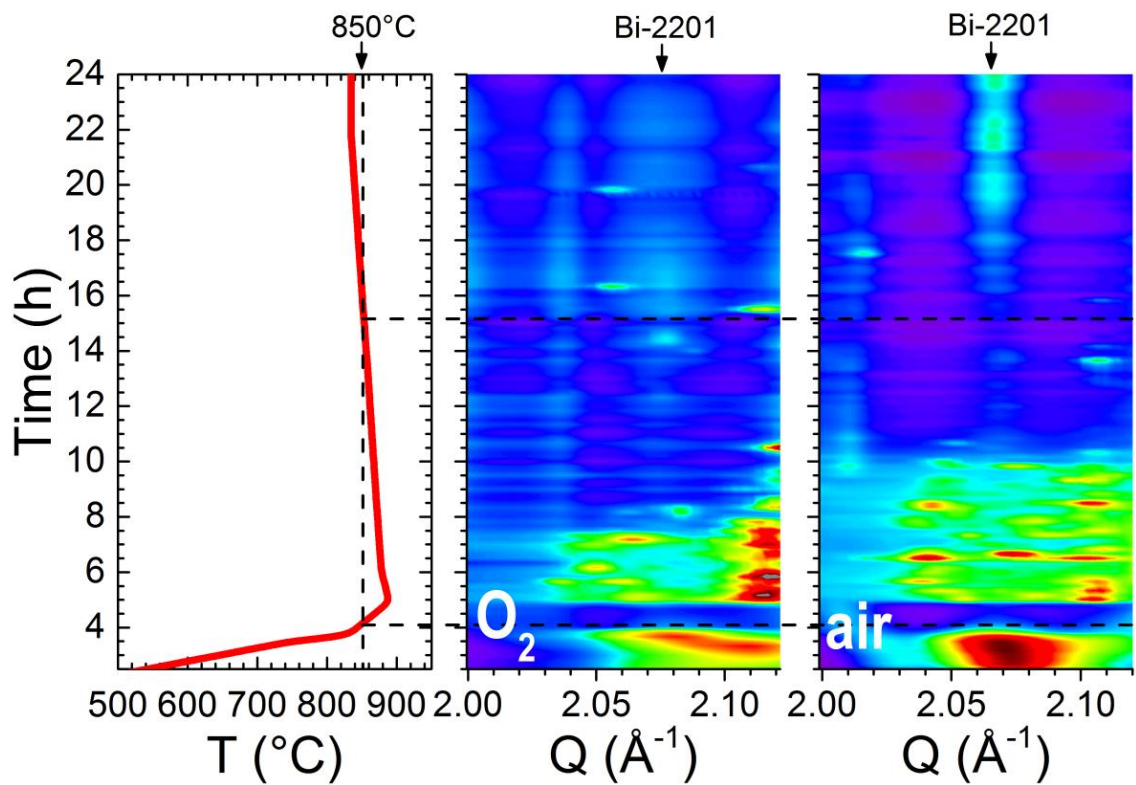

FIG.16. Evolution of the highest intensity 151 diffraction line of the Bi-2201 phase in the $O_2$ and air treated samples (XRPD data). After melting, the presence of this phase is hardly detectable in the $O_2$ treated sample; conversely, it is clearly observed in the air treated sample and its mass percentage increases with time.

The structural data resulting from the fits are reported in Table II; remarkably, the amount of secondary Bi-2201 phase appears to increase with the decrease of the c-axis of the Bi-2212 phase.

Data collected on the $O_2$ treated sample after 24 h (835°C), 29 h (835°C) and 36 h (295°C) during the thermal treatment TT2 are compared in Table III. Seemingly, the prolonged annealing at 835°C produces no significant change of the structural properties of Bi-2212 and even the amount of secondary Bi-2201 remains constant. This conclusion must, however, be taken with caution since it is based on diffraction data of relatively low quality, being influenced by the scattering and absorption of the Ag sheet.

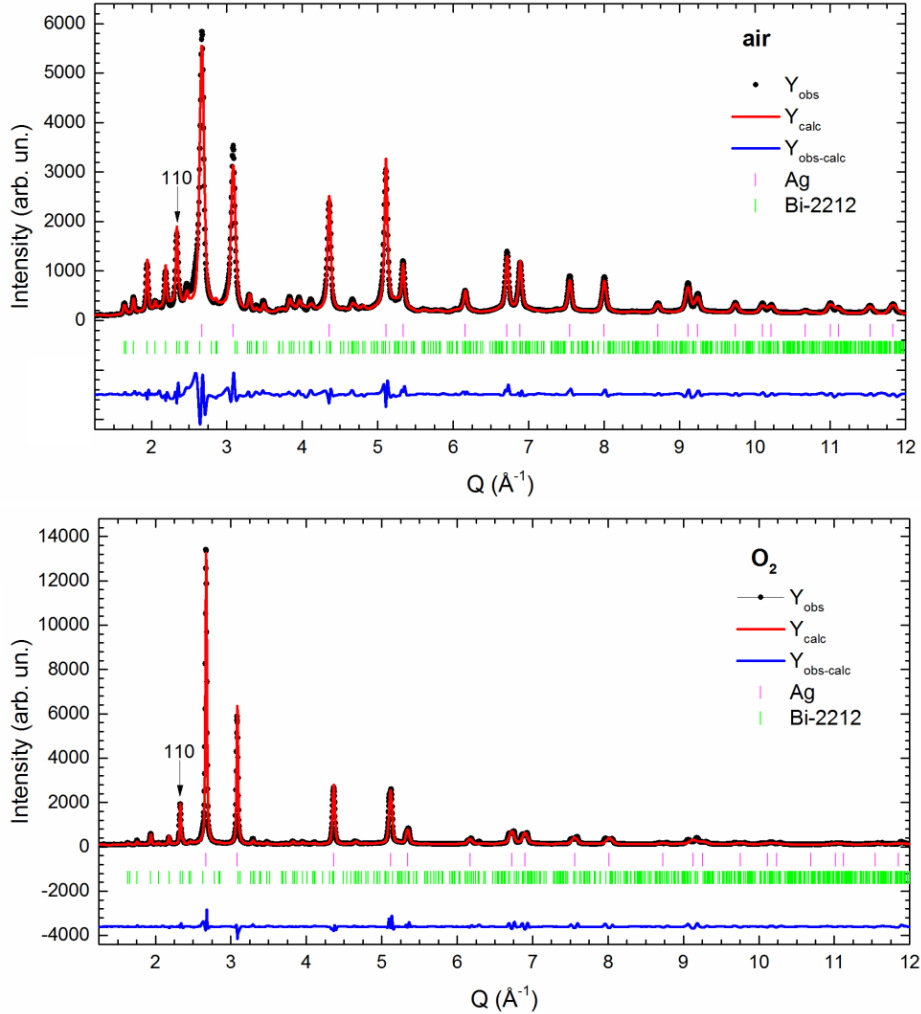

FIG.17. Rietveld refinements obtained for the fully reacted samples (XRD data collected at room temperature); the 110 diffraction line of the Bi-2212 phase marking the texturing of the samples is arrowed; upper and lower vertical bars indicate Bragg reflections for Ag and Bi-2212, respectively.

### G. Microstructural analysis of the silver sheath

The different evolution of the phase assemblages and texturing in the multifilamentary wires treated under air and $O_2$ can be definitely ascribed to the different thermodynamic activity of oxygen within the samples. This phenomenon can be directly probed by the microstructural analysis of the Ag sheath, undergoing microstructural changes during the thermal treatments. Generally speaking, oxygen diffusion initially occurs along the grain boundaries, but as the temperature increases intra-grain interstitial diffusion can be activated and the incorporation of O atoms in the Ag structure consequently induces lattice strains [29]. In particular, oxygen diffusion determines the anisotropic displacement of Ag atoms and the formation of structural defects. As a result, the resulting structural strains selectively broadens the diffraction lines, as evidenced in Figure 18, where the Williamson-Hall plots obtained from the two samples at the beginning (~70°C) and at the end (~888°C) of the heating step are superposed.

|  | air | $O_2$ |
|---|---|---|
| a (Å) | 3.8033(1) | 3.8146(1) |
| c (Å) | 30.6504(6) | 30.7866(7) |
| Bi-2212 (wt%) | 91.22(4.46) | 95.43(5.9) |
| Bi-2201 (wt%) | 8.78(2.0) | 4.47(2.53) |

TABLE II. Cell parameters and weight percentage of $Bi_2Sr_2CuO_{6+x}$ and $Bi_2Sr_2Ca_1Cu_2O_{8+x}$ phases at the end of the thermal treatment TT1 in air and $O_2$ (fully reacted samples; XRD data collected at room temperature).

|  |  | 24h/835°C | 29h/835°C | 36h/295°C |
|---|---|---|---|---|
| Bi-2212 | a (Å) | 3.9301(1) | 3.9297(1) | 3.8994(1) |
|  | c (Å) | 31.635(1) | 31.632(1) | 31.4092(6) |
| Bi-2201 (wt%) |  | 3.0(9) | 3.2(4) | 5.3(5) |

TABLE III. Structural data of Bi-2212 and weight percentage of secondary Bi-2201 phase at different steps during the $O_2$ thermal treatment TT2.

Data collected from the air treated samples display a decrease of the strain along all of the crystallographic directions, in- dicating that the lattice strains produced during the cold working of the wires are elimi- nated by the thermal treatment. Conversely, in the $O_2$ treated sample a notable increase of the lattice strain develops along all the crystallographic direction, thus demonstrating the incorporation of O atoms in the Ag structure, that is an effective diffusion of oxygen throughout the silver sheath. In particular, the microstructural properties observed for this sample are consistent with the results reported by Nagy et al. [29]. These authors stated that oxygen diffusion takes place along [220], thus producing strain broadening of the 220 diffraction line, as well as a similar broadening for the 111 and 311 diffraction lines, whose zone axis at the intersection of the corresponding planes is the same [220]; conversely, the 200 diffraction line exhibits a different behaviour. This phenomenology can be exactly observed by comparing the Williamson-Hall plots of the $O_2$ treated sam- ple. Based on these results, it can be concluded that in the air treated sample oxygen mainly diffuses along grain boundaries and bulk diffusion is essentially hindered; con- versely, the selective lattice strain observed in the $O_2$ treated sample proves the activation of the bulk diffusion, determining a larger thermodynamic activity of the oxygen within the filaments.

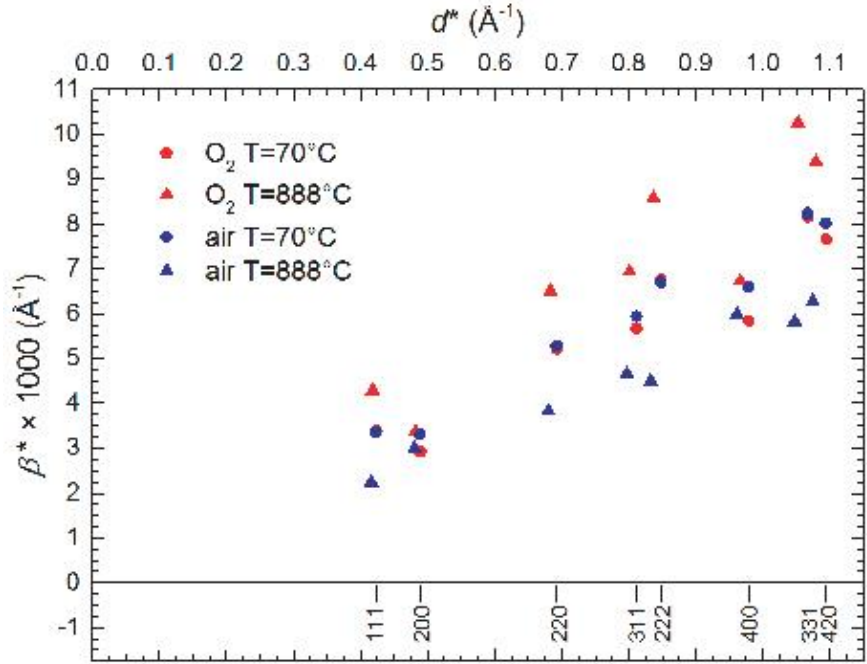

FIG. 18. Superposition of the Williamson-Hall plots obtained from the two samples at the be- ginning (~70°C) and at the end (~888°C) of the heating step; the vertical bars in the lower field indicate the positions of the indexed Bragg peaks.

## IV. Discussion

The aim of the present paper is to give a contribution to the understanding of the pro- cesses actually occurring during the typical heat treatment in the Bi-2212 wire fabrication process. For the first time *in-situ* diffraction data have been collected giving insights into the phase assemblage and microstructural evolution inside a wire having been subjected to exactly the same conditions in which the wire is normally processed for application purposes. The results can thus have a direct repercussion on the production technology of such superconductors, improving their preparation. Furthermore, the observations re- ported here can partly clarify aspects and concerns left unsolved in previous works [14] [15] [16] regarding the phase formation, the evolution of the texture, the role of the oxy- gen on the transport properties and the evolution of the secondary phase Bi-2201. All the analysis was performed on our lab-made wires; however, it is worth to underline that the obtained results are in fully agreement with those observed on state-of-the-art Bi-2212 wires. Looking at the evolution of the phases in Figure 8, we can observe that the Bi-2201 phase starts to grow well below the melting temperature. On the other hand, it is inter- esting to observe that this phase then decomposes again forming novel Bi-2212 phase at 850 °C in both cases, air and $O_2$. If we look at what happens after the melting, during the slow cooling in which the recrystallization occurs (Figure 15), new Bi-2201 phase appears again at about 850 °C, but in this case, its amount is determined by oxygen activity: in air it is about 9%wt while in pure $O_2$ it decreases to about 5%wt (Table II). It is as well inter- esting to notice that this amount is reached for both samples already after a few hours at the plateau at 836 °C, and remains substantially unchanged in the fully reacted samples whose heat treatment includes a plateau at 836 °C for 48 h (Tables II and III).

It is well known that Bi-2201 grains, whose superconducting properties are very poor, represent an obstacle for the supercurrent flowing through Bi-2212 grains. From these results, we get a strong indication about a possible optimisation of the parameters of the "standard" PMP aiming at the reduction of secondary phases and thereby at a $J_c$ enhance- ment in conductors used in practice. A further comment has to be done about the Bi-2201 intragrowth layers. Naderi et al [14] reports on the role of such intragrowths that, when their size is comparable with the coherent length of Bi-2212, can actually act as pinning centres, enhancing the transport properties at high field. We found that their density is approximately estimated as 6.5% in the $O_2$ sample, and almost negligible in the air sam- ple. However, given the difficulty in making such calculation in our experiment, we leave open the issue, pushing it back to a more proper and precise investigation.

The analysis of the texture evolution showed that the deformation process is able to in- duce, similar to the situation in the manufacturing process of Bi-2223 tapes, a preferential orientation of the c axis normal to the RD while no in plane orientation is present after the cold working process or in the first part of the thermal treatment, where the Bi-2212 phase partially recrystallizes. During the "melting" obviously no Bi-2212 crystalline phase is de- tected, but simpler oxides are observed. The diffraction pattern of these phases does not correspond to a powder pattern with circular Debye Scherrer rings circles but appears as bright spots reflecting the formation of a few but large crystal grains. The spots are ran- domly distributed in the azimuthal direction, indicating that no preferential orientation is kept in this stage of the process.

Lowering the temperature, the Bi-2212 phase starts to crystallize (at different temper- atures in $O_2$ atmosphere or air as previously discussed) and nucleates directly with a strong biaxial texture. In particular the $c$-axis of the Bi-2212 grains is narrowly distributed

(FWHM 15) around the normal to the wire plane. A measurement performed on a fully reacted wire, without the furnace but mounted on a Φ stage, allowing rotating the sam- ple around the wire axes, confirmed the cylindrical symmetry by showing no significant dependence of the pattern on the Φ angle. A smaller, but significant orientation (FWHM 30) is observed also for the distribution of the rotation of the grain around their *c*-axis. The observed texture and its values are fully in agreement with the results of Kametani et al [27] but in addition to this work we were able to observe that this texture originates from the nucleation of the Bi-2212 phase after the melt and does not depend on any tex- ture previously induced in the powders. The successive steps in the thermal treatment (slow cooling and plateau) affect the grain growth and the crystalline order of the phase but no effects are observed on the evolution of the texture for the sample treated in $O_2$. The sample reacted in air shows basically the same behaviour. Its diffraction patterns reveals, however, a certain texture evolution during the plateau step and a number of grains showing different orientations are detected. In some cases, a tendency for these grains to align along the "main" orientation in the course of the thermal treatment is ob-served. In general, the degree of the final texture in the air-treated samples is lower than in $O_2$-treated ones indicating that the oxygen activity plays a primary role.

## V. Conclusion

In summary, we have presented an in-depth *in-situ* analysis of the phase and its microstructure evolution in our lab-made Bi-2212 wires taking place during the heat treatment usually applied in the fabrication process. The results in terms of secondary phase detection and grains texture are not only in full agreement with what is reported for state-of-the-art wires, but also add important insights for a further optimization of the process itself. It is now clear when and at which temperature the Bi-2201 phase forms and consequently how to avoid it. As already observed before, the Bi-2212 phase shows an in-plane and an out-of-plane texturing but the here presented results reveal as well that the tex- ture originates from the nucleation of the Bi-2212 grains and does not evolve during the successive steps of the treatment. The role of the oxygen activity has been better clarified with the demonstration that this parameter rules the Bi-2212 texturing. These findings are of great importance for reducing the amount of secondary phases and improving the degree of texture, the two key parameters for obtaining high $J_c$ values.

These results should furthermore promote the development of other HTS materials as isotropic wires through a similar process.

## Acknowledgements

Authors acknowledge Dr. V Diadkin for his kind support and experimental assistance during the data collection at the ID11 beamline of ESRF (proposal MA-3116). Part of this work has been performed in collaboration with A. Ballarino and S. Hopkins from CERN within the Addendum FCC-GOV-CC-0086 of the Memorandum of Under- stading FCC-GOV-CC-0004 for the FCC study.

Finally, this work has been carried out also within the framework of the EUROfusion Consortium and has received funding from the Euratom research and training pro- gramme 2014–2018 under grant agreement No 633053. The views and opinions expressed herein do not necessarily reflect those of the European Commission.

- [1] J. Bascunan, Seungyong Hahn, Youngjae Kim, Jungbin Song, and Y. Iwasa, IEEE Transac- tions on Applied Superconductivity **24**, 1 (2014).

- [2] L. Bottura, G. de Rijk, L. Rossi, and E. Todesco, IEEE Transactions on Applied Superconduc- tivity **22**, 4002008 (2012).

- [3] D. Larbalestier, A. Gurevich, D. M. Feldmann, and A. Polyanskii, Nature **414**, 368 (2001).

- [4] A. Malagoli, F. Kametani, J. Jiang, U. P. Trociewitz, E. E. Hellstrom, and D. C. Larbalestier, Superconductor Science and Technology **24**, 075016 (2011).

- [5] F.Kametani,T.Shen,J.Jiang,C.Scheuerlein,A.Malagoli,M.DiMichiel,Y.Huang,H.Miao, J. A. Parrell, E. E. Hellstrom, and D. C. Larbalestier, Superconductor Science and Technology **24**, 075009 (2011).

- [6] C. Scheuerlein, M. Di Michiel, M. Scheel, J. Jiang, F. Kametani, A. Malagoli, E. E. Hellstrom, and D. C. Larbalestier, Superconductor Science and Technology **24**, 115004 (2011).

- [7] A. Malagoli, P. J. Lee, A. K. Ghosh, C. Scheuerlein, M. Di Michiel, J. Jiang, U. P. Trociewitz, E. E. Hellstrom, and D. C. Larbalestier, Superconductor Science and Technology **26**, 055018 (2013).

- [8] D. C. Larbalestier, J. Jiang, U. P. Trociewitz, F. Kametani, C. Scheuerlein, M. Dalban-Canassy, M. Matras, P. Chen, N. C. Craig, P. J. Lee, and E. E. Hellstrom, Nature Materials **13**, 375 (2014).

- [9] A. Leveratto, V. Braccini, D. Contarino, C. Ferdeghini, and A. Malagoli,

Superconductor Science and Technology **29**, 045005 (2016).

[10] A. Malagoli, V. Braccini, M. Vignolo, X. Chaud, and M. Putti, Superconductor Science and Technology **27**, 055022 (2014).

[11] A. Malagoli, C. Bernini, V. Braccini, G. Romano, M. Putti, X. Chaud, and F. Debray, Super- conductor Science and Technology **26**, 045004 (2013).

[12] T. Shen, J. Jiang, F. Kametani, U. P. Trociewitz, D. C. Larbalestier, J. Schwartz, and E. E. Hellstrom, Superconductor Science and Technology **23**, 025009 (2010).

[13] W. Zhang and E. E. Hellstrom, Superconductor Science and Technology **8**, 430 (1995).

[14] G. Naderi and J. Schwartz, Applied Physics Letters **104**, 152602 (2014).

[15] C. Scheuerlein, J. Andrieux, M. Rikel, J. Kadar, C. Doerrer, M. DiMichiel, A. Ballarino, L. Bot- tura, J. Jiang, F. Kametani, E. Hellstrom, and D. Larbalestier, IEEE Transactions on Applied Superconductivity **26**, 1 (2016).

[16] I. Pallecchi, A. Leveratto, V. Braccini, V. Zunino, and A. Malagoli, Superconductor Science and Technology **30**, 095005 (2017).

[17] R. Flukiger, Y. Huang, F. Marti, M. Dhalle, E. Giannini, R. Passerini, E. Bellingeri, G. Grasso, and J.-C. Grivel, IEEE Transactions on Appiled Superconductivity **9**, 2430 (1999).

[18] E. Giannini, E. Bellingeri, R. Passerini, and R. Flükiger, Physica C: Superconductivity **315**, 185 (1999).

[19] J. Rodríguez-Carvajal, Physica B: Condensed Matter **192**, 55 (1993).

[20] J. I. Langford, D. Louër, E. J. Sonneveld, and J. W. Visser, Powder Diffraction **1**, 211 (1986).

[21] R. A. Young, ed., *The Rietveld method*, International Union of Crystallography monographs on crystallography No. 5 (International Union of Crystallograhy ; Oxford University Press, [Chester, England] : Oxford ; New York, 1993).

[22] J. Torrance, Y. Tokura, S. LaPlaca, T. Huang, R. Savoy, and A. Nazzal, Solid State

Communications **66**, 703 (1988).

[23] P. Bordet, J. Capponi, C. Chaillout, J. Chenavas, A. Hewat, E. Hewat, J. Hodeau, M. Marezio, J. Tholence, and D. Tranqui, Physica C: Superconductivity **153-155**, 623 (1988).

[24] M. Rikel and E. Hellstrom, Physica C: Superconductivity **357-360**, 1081 (2001).

[25] C. C. Luhrs, E. Molins, G. Van Tendeloo, D. Beltrán-Porter, and A. Fuertes, Chemistry of Materials **10**, 1875 (1998).

[26] J. Kadar, C. Scheuerlein, M. O. Rikel, M. Di Michiel, and Y. Huang, Superconductor Science and Technology **29**, 105009 (2016).

[27] F. Kametani, J. Jiang, M. Matras, D. Abraimov, E. E. Hellstrom, and D. C. Larbalestier, Scientific Reports **5** (2015), 10.1038/srep08285.

[28] A. Dellicour, B. Vertruyen, M. Rikel, L. Lutterotti, A. Pautrat, B. Ouladdiaf, and D. Chateigner, Materials **10**, 534 (2017).

[29] A. J. Nagy, G. Mestl, D. Herein, G. Weinberg, E. Kitzelmann, and R. Schlögl, Journal of Catalysis **182**, 417 (1999).